\documentclass[aps]{revtex4}%
\usepackage{amsfonts}
\usepackage{amsmath}
\usepackage{amssymb}
\usepackage{graphicx}%
\begin{document}
\title[Two-dimensional electronic states]{Two-dimensional zero-gap electronic states at a magnetic field}
\author{S.A. Ktitorov}
\affiliation{Ioffe Physico-Technical Institute of the Russian Academy of Sciences}
\author{Yu.V. Petrov}
\affiliation{Ioffe Physico-Technical Institute of the Russian Academy of Sciences}

\begin{abstract}
This work was firstly published in 1986 \cite{we}. No real two-dimensional
object with the zero-gap quasi-relativistic spectrum was known in that time.
Such an object is well known now: this is graphene. That is why we decided to
present it again as a e-print in a slightly modified form.

A density of the two-dimensional zero-gap electronic states at the quantizing
magnetic field in the presence the Gaussian random potential has been
calculated. The problem is reduced to zero-dimensional spinor field theory
using the holomorphic supersymmetric representation. The calculated density of
states in the case of the mass perturbation has a delta function peak in the
Dirac point.This peak smears due to the potential perturbation.

\end{abstract}
\maketitle








\section{Introduction}

A supersymmetry formalism for calculation of the two-dimensional Schroedinger
electronic density of states (EDOS) in the presence of the random potential
was developed in \cite{brezin}. The problem was formulated in terms of the
path integrals. The supersymmetric holomorphic representation \cite{faddeev},
\cite{wegner} was used to project the state onto the lower Landau level. The
problem was reduced to the zero-dimensional field theory after averaging over
the random potential distribution. The EDOS was calculated exactly. This
result was in accord with one obtained in \cite{wegner} for the case of the
Gaussian white noise distribution.

This approach is generalized in this work to the case of the zero-gap
semiconductor with the quasirelativistic spectrum described by the Dirac
equation in the vicinity of the bands crossing point, zero-gap semiconductors
of the first kind (ZGSC-I) \cite{abricos}.

The random background field is created by the distribution short-range
impurities of two kinds: (i) impurities shifting the conduction and valence
bands edges synchronously, i. e. shifting the chemical potential (potential
impurities), and (ii) impurities shifting the bands edges in the opposite
directions, i. e. modulating the gap width (chemical impurities). Such
situation can be realized in IV-VI semiconductors solid solutions, where the
band gap can be made vanishing varying temperature, pressure and composition.
The spatial fluctuations of the composition play a role of the chemical
perturbation, while the random violations of stoichiometry play a role of the
potential perturbation. Small correlation radius of the random potential stems
from the extremely strong lattice polarization in the IV-VI semiconductors:
the Coulomb tail $e^{2}/\left(  \epsilon r\right)  $ can be neglected, when
$\epsilon\symbol{126}400$. In the two-dimensional systems like graphene a
potentials shifting the chemical potential and modulating the mass (gap)
appear because the Kohn-luttenger matrix elements of the short-range impurity
potential are not equal generically for the upper and lower bands.

A character of the EDOS is in accord with \cite{brezin} in the case of Coulomb
impurities. In the case of the chemical impurities there is a delta function
singularity in the centre of the band. Simultaneous effect of the chemical and
Coulomb perturbations smears this singularity, \ when chemical perturbation
dominates. When intensities of these perturbations are equal, EDOS is similar
to one obtained in \cite{brezin}.

\section{Holomorphic representation for ZGSC-I}

Electronic states in ZGSC-I with impurities at the quantizing magnetic field
are described by the two-band Hamiltonian \cite{beer}:%
\begin{equation}
\widehat{\mathcal{H}}\mathcal{=}\widehat{\mathcal{H}}_{0}+\widehat{V}\left(
\mathbf{r}\right)  , \label{ham}%
\end{equation}

\begin{equation}
\widehat{\mathcal{H}}_{0}=s\mathbf{\alpha}\left(  \widehat{\mathbf{p}%
}\mathbf{-}\frac{e}{c}\mathbf{A}\right)  \label{dirac}%
\end{equation}
where $\mathbf{\alpha}$ is the Dirac matrix, s is the Fermi velocity,
$V\left(  \mathbf{r}\right)  $ is the potential of impurities. We choose the
axial gauge for the magnetic field potential: $\mathbf{A}=\frac{1}%
{2}\mathbf{H}\times\mathbf{r}$, $\mathbf{H\parallel e}_{z}.$ The spectrum of
the operator (\ref{dirac}) is well known \cite{johnsonlippmann}. Let us
construct the relativistic holomorphic representation. It is carried out by
functions of the following form:%
\begin{equation}
\Psi_{j}\left(  x,y\right)  =\exp\left[  -\varsigma\overline{\varsigma
}\right]  u_{j}\left(  \varsigma\right)  , \label{holo}%
\end{equation}
where $\varsigma=\left(  x+iy\right)  /2l_{H},$ $\overline{\varsigma}=\left(
x-iy\right)  /2l_{H},$ $l_{H}^{-2}=eH/\hbar c$ is the magnetic length, $j$ is
the spinor index, $u_{j}\left(  \varsigma\right)  $ is a holomorphic function
of the variable $\varsigma$ in the Fock-Bargmann space \cite{bargmann}. We can
make use of the close relation between the holomorphic representation and the
Glauber coherent state representation \cite{carruthers}. The wave functions of
the stationary states with the energy $E$, spin component $\sigma_{z}=1$ and
momentum $k_{z}$ has the following form in the relativistic holomorphic
representation \cite{malkinmanko}:%
\begin{equation}
\Psi_{n,\sigma_{z}=1,k_{z}}\left(  x,y\right)  =\frac{1}{\sqrt{2}}\left[
\begin{array}
[c]{c}%
\left\vert n,\beta\rangle\right. \\
\hbar s\sqrt{n}/\left(  l_{H}\right)  \left\vert n-1,\beta\rangle\right. \\
\left\vert n,\beta\rangle\right. \\
0
\end{array}
\right]  . \label{bispinor}%
\end{equation}
Here $n$ is the Landau band number, $\operatorname{Re}\beta$ and
$\operatorname{Im}\beta$ are the circular motion coordinates. It is seen from
(\ref{4spinor}) that the small components of the bispinor vanish at $n=0.$ The
Dirac bispinor (\ref{4spinor}) reduces into the spinor%
\begin{equation}
\Psi_{0,\sigma_{z}=1,k_{z}}\left(  x,y\right)  =\frac{1}{\sqrt{2}}\left[
\begin{array}
[c]{c}%
1\\
1
\end{array}
\right]  \left\vert 0,\beta\rangle\right.  , \label{spinor}%
\end{equation}
where $\left\vert 0,\beta\rangle\right.  $is an eigenfunction of the operator
$\widehat{b}:$%
\begin{equation}
\widehat{b}\left\vert 0,\beta\rangle\right.  =\beta\left\vert 0,\beta
\rangle\right.  ;\text{ \ \ }\left\vert 0,\beta\rangle\right.  =\left(
\sqrt{2\pi}l_{H}\right)  ^{-1}\exp\left[  -\varsigma\overline{\varsigma}%
+\sqrt{2}\beta\varsigma-\left\vert \beta\right\vert ^{2}/2\right]  ,
\label{eigen}%
\end{equation}

\begin{equation}
\widehat{b}=\frac{1}{\sqrt{2}}\left(  \varsigma+\partial/\partial
\varsigma\right)  ;\text{ \ \ }\widehat{b}^{\dagger}=\frac{1}{\sqrt{2}}\left(
\varsigma-\partial/\partial\varsigma\right)  ;\text{ \ \ }\left[  \widehat
{b},\text{\ }\widehat{b}^{\dagger}\right]  =1. \label{commut}%
\end{equation}
The function $\left\vert 0,\beta\rangle\right.  $ is a generating one for the
states with a definite angular momentum $z-$component quantum number $m.$ Let
us expand $\exp\left[  \left\vert \beta\right\vert ^{2}\right]  \left\vert
0,\beta\rangle\right.  $ to powers of $\beta:$%
\begin{equation}
\exp\left[  \left\vert \beta\right\vert ^{2}\right]  \left\vert 0,\beta
\rangle\right.  =\exp\left(  -\varsigma\overline{\varsigma}\right)  \sum
_{m=0}^{m_{\max}}B_{m}\varsigma^{m}=\exp\left(  -\varsigma\overline{\varsigma
}\right)  u\left(  \varsigma\right)  =\varphi\left(  x,y\right)  ,
\label{expand}%
\end{equation}
where $B_{m}=\frac{1}{\sqrt{2\pi}l_{H}}\beta^{m}/m!,$ $m_{\max}=L^{2}/2\pi
l_{H}^{2}.$ Thus, one can see from (\ref{spinor}) and (\ref{expand}) a
correspondence existing between the coherent and relativistic holomorphic
representations. The Landau zero-band set membership is determined by the
Cauchy-Riemann conditions for $u\left(  \varsigma\right)  :$%
\begin{equation}
\partial u\left(  \varsigma\right)  /\partial\overline{\varsigma}=0,\text{
\ }\partial\text{\ }\overline{u}\left(  \overline{\varsigma}\right)
/\partial\varsigma=0. \label{cauchy}%
\end{equation}
While quantizing we replace the variables $\varsigma$, $\overline{\varsigma}$
with the ladder operators $\widehat{\varsigma}$, $\widehat{\overline
{\varsigma}}=\partial/\partial\varsigma$ acting in the Fock space. Their
commutaters read:%
\begin{equation}
\left[  \widehat{\varsigma},\widehat{\overline{\varsigma}}\right]  =1\text{,
\ \ }\left[  \widehat{\varsigma},\widehat{\varsigma}\right]  =0. \label{comm2}%
\end{equation}

\section{Functional integral representation for Green's functions}

Our goal in this section is to derive a general expression for EDOS for the
quasirelativistic states of the lower Landau band of a two-dimensional system
at the random impurity field. We follow to the approach developed in
\cite{brezin} generalizing it to the quasirelativistic system. A general
expression for EDOS reads:%
\begin{equation}
\rho\left(  E\right)  =-\frac{1}{\pi S}\operatorname{Im}Tr\int
dxdy\left\langle G\left(  E+i0;\mathbf{r=r}^{\prime}\right)  \right\rangle ,
\label{dos}%
\end{equation}
where $G\left(  E+i0;\mathbf{r,r}^{\prime}\right)  $ is a one-particle Green's
function; trace is taken on the spinor indices, $S$\ is the surface area; the
angle brackets indicate averaging on the random potential configurations. The
Green function can be presented by the functional integral:%
\begin{equation}
G_{\alpha j,\alpha^{\prime}j^{\prime}}\left(  E+i0;\mathbf{r,r}^{\prime
}\right)  =-iZ^{-1}\int D\varphi D\varphi^{\ast}\varphi_{\alpha j}\left(
\mathbf{r}\right)  \varphi_{\alpha^{\prime}j^{\prime}}^{\ast}\left(
\mathbf{r}^{\prime}\right)  \exp\left[  i\int dxdy\varphi_{\alpha j}^{\ast
}\left(  \mathbf{r}\right)  \left(  E-\widehat{\mathcal{H}}+i0\right)
\varphi_{\alpha j}\left(  \mathbf{r}\right)  \right]  , \label{green}%
\end{equation}

\begin{equation}
Z=\int D\varphi D\varphi^{\ast}\exp\left[  i\int dxdy\varphi_{\alpha j}^{\ast
}\left(  \mathbf{r}\right)  \left(  E-\widehat{\mathcal{H}}+i0\right)
\varphi_{\alpha j}\left(  \mathbf{r}\right)  \right]  , \label{partition}%
\end{equation}
where $\widehat{\mathcal{H}}$ is determined by (\ref{ham}). The impurity
potential is given by the formulae:%
\begin{equation}
V\left(  \mathbf{r}\right)  =V_{1}\left(  \mathbf{r}\right)  +\sigma_{1}%
V_{2}\left(  \mathbf{r}\right)  ,\text{ \ \ }\left\langle V_{i}\left(
\mathbf{r}\right)  \right\rangle =0,\text{ \ \ }\left\langle V_{i}\left(
\mathbf{r}\right)  V_{j}\left(  \mathbf{r}^{\prime}\right)  \right\rangle
=\lambda_{i}\delta_{ij}\delta\left(  \mathbf{r-r}^{\prime}\right)  .
\label{impur}%
\end{equation}
Here $V_{1}\left(  \mathbf{r}\right)  $ is a screened Coulomb potential,
$V_{2}\left(  \mathbf{r}\right)  $ is a mass (gap) modulating perturbation;
$\sigma_{1}=\left[
\begin{array}
[c]{cc}%
0 & 1\\
1 & 0
\end{array}
\right]  $ is the Pauli matrix.

Let us define a supervector belonging to the zero Landau level subspace:%
\begin{equation}
\Phi=\left[
\begin{array}
[c]{c}%
\varphi\\
\psi
\end{array}
\right]  ,\text{ \ \ }\psi_{j}=\exp\left(  -\varsigma\overline{\varsigma
}\right)  v_{j}\left(  \varsigma\right)  ,\text{ \ \ }\overline{\Phi}=\left(
\varphi^{\ast},\overline{\psi}\right)  , \label{supervector}%
\end{equation}
where $v_{j}\left(  \varsigma\right)  $ is the Grassmann variable. Then the
averaged Green function reads in terms of the supervectors:%
\begin{equation}
G_{ij}=-i\int D\Phi D\overline{\Phi}u_{i}\left(  \varsigma\right)  u_{j}%
^{\ast}\left(  \overline{\varsigma}\right)  \exp\left(  -\varsigma
\overline{\varsigma}\right)  \exp\left(  iS_{hol}\right)  , \label{supergreen}%
\end{equation}
where the effective action reads%
\begin{equation}
S_{hol}=E\int d\varsigma d\overline{\varsigma}\exp\left(  -\varsigma
\overline{\varsigma}/2\right)  \overline{\Phi}\Phi+\int d\varsigma
d\overline{\varsigma}\exp\left(  -\varsigma\overline{\varsigma}\right)
\left[  \lambda_{1}\left(  \overline{\Phi}\Phi\right)  +\lambda_{2}\left(
\overline{\Phi}\sigma_{1}\Phi\right)  ^{2}\right]  . \label{action}%
\end{equation}
Here $\lambda_{1}$ and $\lambda_{2}$ are intensities of the impurity
correlators (\ref{impur}). It is seen from (\ref{action}) that varying of the
free action with respect to $\overline{\Phi}$ gives the equation for the
Landau gapless zero-band; in the two-dimensional case it is a zero-energy
state
\begin{equation}
\mathcal{H}_{0}\Phi=0. \label{zero}%
\end{equation}

Following \cite{brezin}, we introduce the holomorphic superfields:%
\begin{equation}
\chi_{i}=u_{i}\left(  \varsigma\right)  +\frac{1}{\sqrt{2}}\theta_{i}%
v_{i}\left(  \varsigma\right)  ,\text{ \ \ }\overline{\chi}_{i}=u_{i}^{\ast
}\left(  \overline{\varsigma}\right)  +\frac{1}{\sqrt{2}}\overline{v}%
_{i}\left(  \varsigma\right)  \overline{\theta}_{i}, \label{superfield}%
\end{equation}
where $\theta_{i},$ $\overline{\theta}_{i}$ are Grassmannian algebra
generators. The effective action presented in terms of the superfields
$\chi_{i}$ is supersymmetric similarly to the Schroedinger case \cite{brezin},
i. e. it is invariant under the magnetic translation group $\left(
\text{translation }\times\text{ gauge transformation}\right)  $ for the Dirac
electron \cite{malkinmanko} and the superspace rotation group:%
\begin{align}
S  &  =i2\pi l_{H}^{2}\int d\theta d\overline{\theta}d\varsigma d\overline
{\varsigma}\exp\left[  -\frac{1}{2}\left(  \varsigma\overline{\varsigma
}+\theta\overline{\theta}\right)  \right]  \overline{\chi}_{i}\chi
_{i}-\nonumber\\
&  2\pi l_{H}^{2}\int d\theta d\overline{\theta}d\varsigma d\overline
{\varsigma}\exp\left[  -\frac{1}{2}\left(  \varsigma\overline{\varsigma
}+\theta\overline{\theta}\right)  \right]  \left[  \lambda_{1}\left(
\overline{\chi}_{i}\chi_{i}\right)  ^{2}+\lambda_{21}\left(  \overline{\chi
}_{i}\sigma_{1}\chi_{i}\right)  ^{2}\right]  , \label{superspaceaction}%
\end{align}

\begin{align}
\chi_{i}\left(  z,\theta\right)   &  =\chi_{i}\left(  z-a,\theta
-\omega\right)  \exp\left\{  -\frac{1}{2}\left[  za^{\ast}+\theta
\overline{\omega}+\frac{1}{2}\left(  \left\vert a\right\vert ^{2}+\left\vert
\omega\right\vert ^{2}\right)  \right]  \right\}  ,\nonumber\\
\delta z  &  =\overline{\omega}\theta,\text{ \ \ }\delta\theta=\omega z.
\label{susytrans}%
\end{align}

Substituting (\ref{superspaceaction}) into (\ref{green}) and calculating the
Green function perturbatively we can see that the superspace Gaussian integral
equals unity in all orders, while the corrections to the bare Green function
are nothing but the zero-dimensional spinor $\lambda\varphi^{4}$ field theory
symmetry coefficients. The total Green function can be written as a ratio of
ordinary Rienannian integrals; the resulting formula differs from the derived
in \cite{brezin} one only by the spinor structure presence.

\bigskip

\section{EDOS calculation}

A general formula for calculation of the EDOS for ZGSC-I at the quantizing
magnetic field in the presence of the random imurity field reads:%
\begin{equation}
\rho\left(  E\right)  =\frac{1}{2\pi S}\operatorname{Im}\frac{\partial
}{\partial\epsilon}\log\int_{0}^{\infty}dxdy\exp\left[  i\alpha\epsilon\left(
x+y\right)  -\alpha\lambda_{1}\left(  x+y\right)  ^{2}-\alpha\lambda
_{2}(x-y)^{2}\right]  , \label{dosintegral}%
\end{equation}
where $\alpha=2\pi l_{H}^{2},$ $x=\left\vert u_{1}\right\vert ^{2},$
$y=\left\vert u_{2}\right\vert ^{2},$ $u_{i}$ are spinor components,
$\epsilon=E+i0.$ Notice that the total number of unperturbed states per the
unit area resulting from (\ref{dosintegral}) is twice larger, than in the
one-band case $\int_{-\infty}^{\infty}dE\rho_{0H}\left(  E\right)  =1/2\pi
l_{H}^{2}.$ We will show below by the direct calculation that a presence of
the chemical impurities modulating the mass (gap) leads to the singularity of
EDOS at $E=0.$ Assuming $\lambda_{1}<<\lambda_{2}$ we can one of the
integrals:%
\begin{align}
\rho\left(  E\right)   &  =\frac{1}{\pi S}\operatorname{Im}\frac{\partial
}{\partial\epsilon}\log\left[  -\sqrt{\pi/4\alpha\lambda_{1}}\exp\left(
-\alpha\epsilon^{2}/4\lambda_{1}\right)  \operatorname{erf}c\left(
-i\frac{\epsilon}{2}\sqrt{\alpha/\lambda_{1}}\right)  \right.  +\nonumber\\
&  \left.  \int_{0}^{\infty}dx\exp\left(  2i\alpha\epsilon x\right)
\operatorname{erf}c\left(  x\sqrt{\alpha\lambda_{2}}+i\frac{\epsilon}{2}%
\sqrt{\alpha/\lambda_{2}}\right)  \right]  \label{firstint}%
\end{align}
We have neglected the term $4\alpha\lambda_{1}x^{2}$ in the exponent since
convergence of the integral at the upper limit is guarantied by the
complementary error function ($\operatorname{erf}c(x)\rightarrow0$ at
$x\rightarrow\infty$ within the domain $\left\vert \arg z\right\vert <\pi/4).$
The integral in (\ref{firstint}) can be calculated exactly \cite{prud}:%
\begin{align}
\rho\left(  E\right)   &  =\sqrt{2/\lambda_{1}}\frac{1}{2\pi^{2}l_{H}}%
\frac{\left(  \pi/4\right)  \exp\left(  -\varepsilon^{2}/\lambda^{2}\right)
+\exp\left(  \varepsilon^{2}\right)  \left[  F\left(  \varepsilon
/\lambda\right)  F\left(  \varepsilon\right)  \left(  1-\lambda^{2}%
/2\varepsilon^{2}\right)  -\frac{\lambda}{2\varepsilon}\left(  F\left(
\varepsilon\right)  -F\left(  \varepsilon/\lambda\right)  \right)  \right]
}{\left[  \left(  \sqrt{\pi}/2\right)  \exp\left(  -\varepsilon^{2}%
/\lambda^{2}\right)  -\frac{\lambda}{\sqrt{\pi}\varepsilon}\exp\left(
\varepsilon^{2}\right)  F\left(  \varepsilon\right)  \right]  ^{2}+\left[
F\left(  \varepsilon\right)  \right]  ^{2}}+\nonumber\\
&  \frac{1}{2\pi l_{H}^{2}}\delta\left(  E\right)  , \label{integrated}%
\end{align}
where $\varepsilon=\frac{E\sqrt{\alpha/\lambda_{2}}}{2},$ \ \ $\lambda
=\sqrt{\lambda_{1}/\lambda_{2}},$ \ $F\left(  x\right)  =\exp\left(
-x^{2}\right)  \int_{0}^{x}dt\exp t^{2}.$

Let us consider some limiting cases. If $\lambda_{2}=0,$ i. e. chemical
impurities modulating the mass (gap) are absent, we can simplify
(\ref{integrated}):%
\begin{equation}
\rho\left(  E\right)  =\sqrt{2/\lambda_{1}}\frac{1}{4\pi^{2}l_{H}}\frac
{\exp\left(  \eta^{2}\right)  }{\eta^{2}+\left[  \frac{1}{\sqrt{\pi}}%
\exp\left(  \eta^{2}\right)  \frac{d}{d\eta}F\left(  \eta\right)  \right]
^{2}}, \label{coulomb}%
\end{equation}
where $\eta=\frac{E}{2}\sqrt{\alpha/\lambda_{1}}.$

Taking account of the delta-correlated chemical impurities only ($\lambda
_{1}=0,$ \ \ $\lambda_{2}\neq0)$ gives a delta function peak at $E=0$ on the
smooth background:%
\begin{equation}
\rho\left(  E\right)  =\sqrt{\pi/2\lambda_{2}}\frac{1}{\pi^{2}l_{H}}\left\{
\frac{\pi}{2}\delta\left(  \varepsilon\right)  +\frac{\exp\left(
\varepsilon^{2}\right)  /\sqrt{\pi}}{1+\left[  \frac{2}{\sqrt{\pi}}\exp\left(
\varepsilon^{2}\right)  F\left(  \varepsilon\right)  \right]  ^{2}}\right\}  .
\label{rhochem}%
\end{equation}

Notice that in the case of $\lambda_{1}=\lambda_{2}=w/8$ we obtain EDOS
similar to obtained in (\ref{brezin}), but with the additional facto 2:%
\begin{align}
\rho\left(  E\right)   &  =\sqrt{2\pi/w}\frac{2}{\pi^{2}l_{H}}\frac
{\exp\left(  \nu^{2}\right)  }{1+\left[  \frac{2}{\sqrt{\pi}}\int_{0}^{\nu
}dt\exp\left(  t^{2}\right)  \right]  ^{2}},\label{equallamdas}\\
\nu &  =E\sqrt{\alpha/w}.\nonumber
\end{align}

In the limit of the Coulomb impurity correlator low intensity at $4\lambda
_{1}/\alpha\symbol{126}E^{2}<<4\lambda_{2}/\alpha$ and assuming $E\rightarrow
0,$we obtain from (\ref{integrated}):
\begin{equation}
\rho\left(  E\right)  =\sqrt{2\pi/\lambda_{1}}\frac{1}{8\pi^{2}l_{H}}%
\frac{\sqrt{\pi}\exp\left(  \varepsilon^{2}\right)  }{F^{2}\left(
\varepsilon/\lambda\right)  +\left[  \frac{\sqrt{\pi}}{2}\exp\left(
-\varepsilon^{2}/\lambda^{2}\right)  \right]  ^{2}}. \label{small}%
\end{equation}

It is seen that the peak width is proportional to $\sqrt{2\lambda_{1}/\alpha
}.$ When $E^{2}>>4\lambda_{2}/\alpha>>4\lambda_{1}/\alpha,$ we have a resuld
similar to obtained in (\ref{brezin}) and (\ref{wegner}) for the one-band
semiconductor:%
\begin{equation}
\rho\left(  E\right)  =\sqrt{2}l_{H}E^{2}\frac{\exp\left(  -\pi l_{H}^{2}%
E^{2}/2\lambda^{2}\right)  }{4\lambda_{2}^{3/2}}. \label{highenergy}%
\end{equation}
Such asymptotic was obtained in \cite{affl} by the semi-classical quantization approach.

Narrow peak appear in the strongly irregular semiconductors with the Lorentz
distribution. The effective action takes the form $S=-\lambda x.$This gives a
generalization of the Lloyd model \cite{lloyd} to the zero-gap semiconductor.
The EDOS takes the form%
\begin{equation}
\rho\left(  E\right)  =\frac{1}{2\pi^{2}l_{H}^{2}}\left[  \frac{\Lambda_{+}%
}{\Lambda_{+}^{2}+E^{2}}+\frac{\Lambda_{-}}{\Lambda_{-}^{2}+E^{2}}\right]  ,
\label{lloyd}%
\end{equation}

where $\Lambda_{\pm}=\lambda_{1}\pm\lambda_{2}.$ When $\Lambda_{-}%
\rightarrow0$ (close values of intensities), the peak width tends to zeo.

\bigskip

\section{Discussion}

In the absence of dynamic interactio the action (\ref{superspaceaction}) is
invariant with respect to the supersymmetry transformations. It is doubly
degenerate (apart from the usual Landau degeneracy in the magnetic field);
$E=0.$ Impurities do not violate the supersymmetry that results in appearing
of the delta-peak. Possible physical realization: surface states, states in
the supersymmetric interface in the heterojunction of mutually inverted
narrow-gap semiconductors \cite{volkov}, and, now the most interesting, graphene.

\end{document}